\documentclass{article}

\usepackage{PRIMEarxiv}
\usepackage{caption}
\usepackage[utf8]{inputenc} 
\usepackage[T1]{fontenc}    
\usepackage{hyperref}       
\usepackage{url}            
\usepackage{booktabs}       
\usepackage{amsfonts}       
\usepackage{nicefrac}       
\usepackage{placeins}
\usepackage{microtype}      
\usepackage{lipsum}
\usepackage{fancyhdr}       
\usepackage{graphicx}       
\usepackage{amssymb}        
\usepackage{amsmath}       
\graphicspath{{figures/}}     

\newcommand\Tau{\mathcal{T}}

\usepackage{placeins}
\usepackage{soul,color}
\title{Materials Property Prediction with Uncertainty Quantification: A Benchmark Study
\thanks{\textit{\underline{Citation}}: 
\textbf{D.Varivoda. title. Pages.... DOI:000000/11111.}} 
}

\author{
  Daniel Varivoda \\
Khoury College of Computer Science, Northeastern University, Boston, MA 02115\\
Department of Computer Science and Engineering, University of South Carolina, Columbia, SC 29201 \\
  \And
  Rongzhi Dong, Sadman Sadeed Omee \\
 Department of Computer Science and Engineering\\
  University of South Carolina\\
  Columbia, SC 29201 \\  
   \And
 Jianjun Hu *\\
 Department of Computer Science and Engineering\\
  University of South Carolina\\
  Columbia, SC 29201 \\
  \texttt{jianjunh@cse.sc.edu} \\
}

\begin{document}
\maketitle

\begin{abstract}

Uncertainty quantification (UQ) has increasing importance in building robust high-performance and generalizable materials property prediction models. It can also be used in active learning to train better models by focusing on getting new training data from uncertain regions. There are several categories of UQ methods each considering different types of uncertainty sources. Here we conduct a comprehensive evaluation on the UQ methods for graph neural network based materials property prediction and evaluate how they truly reflect the uncertainty that we want in error bound estimation or active learning. Our experimental results over four crystal materials datasets (including formation energy, adsorption energy, total energy, and band gap properties) show that the popular ensemble methods for uncertainty estimation is NOT the best choice for UQ in materials property prediction. For the convenience of the community, all the source code and data sets can be accessed freely at \url{https://github.com/usccolumbia/materialsUQ}.

\end{abstract}


\section{Introduction}


Historically, the materials discovery process has been a costly endeavour due to the amount of time and resources required for experiments. Due to the enormous amount of possible material compositions, traditional trial-and-error methods can take decades from initial research to first use. Two widely adopted conventional methods, computational simulation and experiments, have some major limitations which make them less than ideal for the acceleration of materials discovery. Generally speaking, experimental investigation is an intuitive method of materials research. However, it tends to require a high material and equipment cost along with a great amount of time. On the other hand, computational simulation, with methods like electronic structure calculation based on density functional theory \cite{schleder2019dft, korzdorfer2014organic, maurer2019advances, kohn1965self, hohenberg1964inhomogeneous}, molecular dynamics \cite{ding2009molecular}, or the phase-field method \cite{steinbach2009phase, fallah2012phase}, exploits existing theory for analysis through the use of computer programs. These simulation methods are widely used in screening existing materials repositories for discovering materials with novel functions \cite{zhao2021screening,zhang2019high}. While computational simulation tends to speed up this process greatly, it can be very computationally expensive and the models are usually specific to the given materials system.

Benefiting from the large amounts of data generated from previous experiments and calculation methods, along with continued advancements in algorithms, machine learning (ML) has begun to revolutionize and greatly accelerate the materials discovery process \cite{hill2016materials, himanen2019data, agrawal2016perspective, calderon2015aflow, hachmann2011harvard, jain2013commentary}. Compared to the previously mentioned classic computational methods such as Density Functional Theory (DFT), ML methods can use a data-driven approach to train materials property prediction models without any mechanistic understanding of the structure-property relationships. By utilizing ML models to predict target material properties, propose potential candidates, and validate with experiments, researchers can greatly increase the efficiency of the materials discovery process. Recently, a large number of machine learning models have been proposed for predicting a variety materials properties based on composition \cite{dan2020computational,jha2021enabling} or structures \cite{xie2018crystal,omee2022scalable} ranging from formation energy, band gaps, elastic constants (bulk and shear modulus) \cite{wang2019machine,zhao2020predicting}, hardness \cite{mazhnik2020application,al2021high}, piezoelectric coefficients \cite{hu2022piezoelectric}, thermoelectric coefficients \cite{han2021machine}, thermal conductivity \cite{loftis2020lattice,ouyang2021machine}, electrode voltage \cite{louis2022accurate}, superconductor critical temperature \cite{belli2021strong,xie2022machine}, XRD spectrum \cite{dong2022deepxrd}, DFT-calculated properties \cite{gupta2021cross}, etc. These ML models have significantly accelerated the discovery of new functional materials. However, most of these models or studies do not provide uncertainty estimation for their predictions except an overall performance measure such as RMSE, $R^2$, or MAE, which neglects the difference of the prediction quality of the query samples located in different areas of the design space. Despite availability of a large number of uncertainty quantification methods including those specifically developed for deep learning \cite{abdar2021review}, there have been limited studies on how these methods work within the area of materials informatics.

\paragraph{Role of UQ in materials informatics}
Machine learning based materials property prediction models have inherent uncertainty due to a variety of sources such as the noise and bias in the input data, incomplete coverage of the domain, and imperfect model of the relationship between structure and properties \cite{nigam2021assigning}. Quantifying the uncertainty or assigning the confidence to the predicted properties for a given sample has many benefits in ML guided materials discovery. First, in material discovery applications where model predictions are used to direct experimental design, UQ can be used to estimate unanticipated imprecision and avoid wasting valuable time and resources. Second, materials discovery projects usually aim to find exotic/novel/outlier materials with exceptional properties, which usually correspond to the out-of-domain samples. The UQ of these outlier samples can guide the prioritizing expensive and time-consuming experimentation \cite{hirschfeld2020uncertainty}. Uncertainty estimation of predicted properties of unknown samples can also be used to improve the explainability of the neural models, which are becoming increasingly standard yet are challenging to interpret
their predictions. UQ results can also help to evaluate the robustness, out-of-domain applicability, and possible failure modes of ML models. Finally, the uncertainty estimation of these models can also be used naturally with active machine learning algorithms for efficient exploration of the design space \cite{xin2021active}. 



\paragraph{Survey of uncertainty estimation methods in machine learning}
In the past years, UQ methods played a pivotal role in reducing the impact of uncertainties during both optimization and decision making processes\cite{abdar2021review}. UQ is also an important component of molecular property prediction, especially where prediction model is then be used to direct experimental design and unanticipated imprecision wastes valuable time and resources\cite{hirschfeld2020uncertainty},
These uncertainty estimates are instrumental for determining which materials to screen next. Tran et al\cite{tran2020methods} use a suite of performance metrics derived from the machine learning community to judge the quality of uncertainty estimates, and find that the convolutional neural network model achieve best performance. Blundell et al\cite{pmlr-v37-blundell15} introduce a new, efficient algorithm for learning a probability distribution on the weights of a neural network. By modeling Bayesian neural network using the theory of subjective logic, Sensoy et al \cite{NEURIPS2018_a981f2b7} improves detection of out-of-distribution queries and endurance against adversarial perturbations. Amini et al\cite{amini2020deep} propose a non-Bayesian neural network model to estimate a continuous target as well as its associated evidence in order to learn both aleatoric and epistemic uncertainty. Zhang et al\cite{zhang_conf_2021} investigate the combination of deep learning-based predictors with the conformal prediction framework to generate highly predictive models with well-defined uncertainties. As distribution UQ model is unsuitable for high-risk settings, Angelopoulos et al\cite{gentle_conf} introduce the distribution-free uncertainty quantification (distribution-free UQ) for creating statistically rigorous confidence intervals/sets without distributional assumption.
Distribution-free UQ also plays an important role in iterative screening. Svensson et al\cite{svensson_conf_2018} use a conformal predictor to maximise gain in iterative screening. Ensemble methods are learning algorithms that construct a set of classifiers and then classify new data points by taking a (weighted) vote of their predictions.\cite{diet_ens} The original ensemble method is Bayesian averaging, and Bayesian neural networks are currently the state-of-the-art for estimating predictive uncertainty. However, this Bayesian method requires large amount of computation cost. Lakshminarayanan et al\cite{lakshminarayanan2017simple} propose an alternative to Bayesian NNs which is parallelizable, requires much less hyperparameter tuning, and yields high quality uncertainty estimates. Drop-out based Bayesian uncertainty measures captures uncertainty better than straightforward alternatives in diagnosing diabetic retinopathy\cite{leibig2017leveraging}
. Isidro et al\cite{cortes2018deep} present a versatile Deep Confidence framework to compute valid and efficient confidence intervals for individual predictions using the deep learning technique Snapshot Ensembling and conformal prediction. This framework can be applied to any deep learning-based application at no extra computational cost. However, this model require more sophisticated learning rate annealing schedule. Wen et al\cite{wen2020uncertainty} propose a class of Dropout Uncertainty Neural Network (DUNN) potentials that can deal with interatomic potentials uncertainty estimate. DUNN potential can provide a rigorous Bayesian estimate for the uncertainty for any predicted property. While there are many UQ methods arising from different application areas, only one recent work \cite{tran2020methods} has conducted benchmark evaluation of six UQ methods for materials property prediction including ensemble neural networks, Bayesian neural networks, dropout neural networks, error/residual prediction networks, Gaussian process regressors with neural network mean ($GPNN_\mu$), and Convolution-Fed Gaussian Process (CFGP). However, the $GPNN_\mu$ is trained with composition features rather than crystal structures. In addition, this early evaluation does not include the most recent UQ methods such as the deep evidential regressor \cite{amini2020deep} and delta-metric method \cite{korolev2022universal}. In addition, direct prediction of errors/residuals has the issue of accuracy.


In this paper, we focus on UQ for Graphical Neural Networks (GNNs)\cite{scarselli2008graph} for materials property prediction, which have recently been at the forefront of materials science research due to their ability to utilize the complex topological relationships among atoms with state-of-the-art performance for molecule and materials property prediction \cite{gnn_base, omee2022scalable, matDL_pap, schutt2017schnet, schwarzer2019learning, chen2019graph,wieder2020compact}. Unlike previous ML models, which aim to reduce graph representations to lower dimensional forms and thus could lose some important information, many GNNs purposefully extract higher level representations. There are already many types of GNNs which, as categorized by Wu et al.\cite{wugnnsurvey}, consisting of Recurrent Graph Neural Networks, Convolutional Graph Neural Networks, Graph Autoencoders, and Spatial–Temporal Graph Neural Networks, with improved networks constantly being manufactured in this relatively new subset of ML.

While GNNs and other NN architectures for quantitative structure-activity relationship (QSAR) modeling have greatly accelerated the expensive and time-consuming process of experimental materials discovery, they are also very opaque, making it difficult to interpret their predictions. The 'black box' nature of these architectures makes it challenging to understand their robustness, out-of-domain applicability, and possible areas of failure \cite{hirschfeld2020uncertainty}. Much focus has been on increasing the overall accuracy of various NN-based QSAR models. However, understanding where a model is confident is also extremely important in materials discovery. Materials property prediction models with uncertainty estimation have the potential to expedite decision-making for experimental validation by considering the balance of risks and benefits for a given candidate material. Recently there has been a significant amount of effort to study UQ for NNs in an attempt to alleviate these issues stemming from their opacity in both QSAR and non-QSAR-related contexts \cite{tran2020methods, amini2020deep, zhang_conf_2021,svensson_conf_2018, lakshminarayanan2017simple, wen2020uncertainty}. Though many different methods have been developed to estimate uncertainty, the differences in data sets, evaluation criteria, hyper-parameter selection, and the models utilized between experiments have made it difficult to compare them. As such, our goal in this paper is to create an objective comparison between some popular UQ techniques and showcase their applications in a materials science setting. Moreover, due to the rapidly evolving nature of NN QSAR research in the materials science space, new models with state-of-the-art accuracy are being developed at a relatively fast pace. With this in mind, we chose to utilize techniques that can be easily applied to any model with a relatively low difficulty, allowing these methods to be replicated with novel models and data sets. Specific methods like UQ through the use of Bayesian networks are not included in our evaluation as while they can provide confidence by varying the posterior on the network weights and then provide estimations of uncertainty, they are also impossible to implement in the context of a non-Bayesian framework such as our graph neural network. In this work, we explore and evaluate four different UQ techniques for GNN based materials property prediction. In the most basic sense, prediction error and model uncertainty should be strongly correlated as a higher error, in theory, should produce a larger uncertainty margin. To test for this, we utilize a basic regression comparing uncertainty and confidence interval size. We also evaluate whether the confidence intervals reflected in our error outputs reflect the distribution of observed model errors, which can be help understand whether predictions are likely to over- or under-confident.

\begin{figure}[ht]
  \centering
  \includegraphics[width=.95\linewidth]{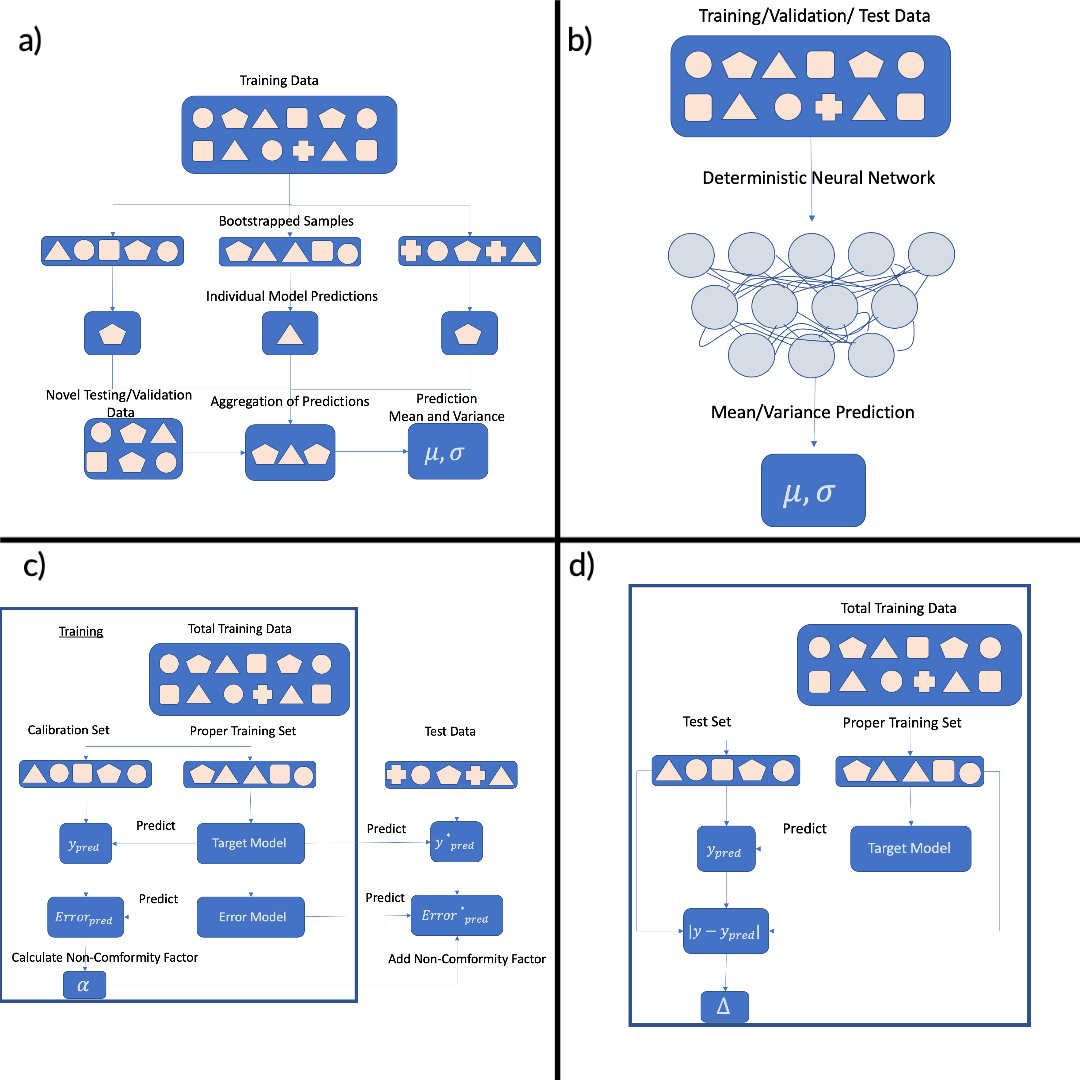}
  \caption{Illustration of the basic architectures for the three UQ methods evaluated in this study. The methods shown from left to right: (a) Ensemble/Bootstrap Ensemble training and prediction, (b) Evidential (Deterministic) model training and prediction, (c) Inductive Conformal model training and prediction, and (d) Delta Method model training and prediction}
  \label{fig:architecture}
\end{figure}

\section{Methods}
\label{sec:headings}
We evaluate four primary techniques for uncertainty estimation of our graph neural network based materials prediction models: Bootstrap Ensemble, Inductive Conformal Prediction, Evidential Learning, and the Delta metric (Figure \ref{fig:architecture}. Each of these techniques was chosen to represent a different overarching strategy for GNN uncertainty prediction. The MEGNet model~\cite{chen2019graph} and MatDeepLearn framework~\cite{matDL_pap} was utilized as the basis framework for this study due to their strong performance. A single model was used as the basis for the four methods implemented in this study to help alleviate issues stemming from inconsistent prediction accuracy due to model variation, allowing for a stronger comparison of uncertainty between methods. Due to the nature of the Evidential method, the MEGNet model could not be used in its unaltered state so only the output layers of the network were changed to maintain model integrity.

\subsection{Bootstrap Aggregation Ensemble}

Ensemble methods \cite{diet_ens} are one of the most widely used uncertainty prediction techniques in machine learning, and it has been shown that ensemble models can create models with a greater accuracy than any one individual model in the set. Rather than training a single model $M$, ensemble based methods instead train a set of models $\xi = \{M_{1}, ..., M_{i}\}$ where each $M_{i}$ represents a distinct model in the ensemble. As such, for any input $x$, the prediction of the ensemble can be defined as the algebraic mean of all ensemble member outputs
\begin{equation}
    \tilde{M}(x) = \sum^{n}_{M \in \xi} \frac{M(x)}{n}
\end{equation}
To approximate uncertainty, we can also use this combination of predictions to approximate a frequency distribution and thus, as a measure of uncertainty, the variance for the model, $V$:
\begin{equation}
    {V}(x) = \sum^{n}_{M \in \xi} \frac{(\tilde{M}(x) - M(x))^{2}}{n}
\end{equation}
There are a few ways ensemble learning can introduce variance in order to create a distribution of responses. In traditional ensembling, each $M_{i} \in \xi$ is trained independently on the same set of training data, $\tau$, but is initialized with different starting weights, with the belief that regions sparsely covered by the training data are more heavily influenced by the initialization. In our experiment we use the bootstrap aggregation method to introduce this variance. In the bootstrap method utilized for this experiment each model $M_{i} \in \xi$ is trained independently as before, however now each model is now trained on a unique subset on a unique subset of the data, $\tau_{i} \in \tau$. This subset is a set of data the same size as the original data set, created by randomly sampling with replacement from the original training set. Sampling with replacement ensures each bootstrap is independent from its peers, as it does not depend on previous chosen samples when sampling. By training each ensemble model on an unique subset of data, the model gains information regarding the density of different features within the input so features which are sparse in the data set should produce high variation between model outputs. The use of bootstrapping severely decreases the amount of data required for training, as in traditional ensemble techniques the size of of the overall training data set needs increase for each new model trained, making it less feasible in areas where data is sparse. The largest drawback of ensemble based methods is the increased resources it utilizes, which varies with the size of the ensemble and the models utilized. For testing purposes, we fix the size of the ensemble to 10.

\subsection{Inductive Conformal Prediction}

Conformal Prediction \cite{eklund_conf_2015} is a newer set of methods for the generation of uncertainty that has recently been gaining attention since they allow the user to set intuitive confidence intervals (e.g. 10\%, 20\%, etc.) and have a strong mathematical backing \cite{eklund_conf_2015,svensson_conf_2018, zhang_conf_2021}. In the most basic sense, these methods involve one model that is trained to predict a property and then a calibration set is utilized to create an uncertainty measure for those predictions. This calibration set can be thought of as a second training set, where the original training set is used to train the model and the calibration set is used to train the confidence. Angelopoulos and Bates' excellent write up \cite{gentle_conf} on the subject outlines the conformal prediction process for a general (categorical) input $x$ and output $y$ as: 
\begin{enumerate}
  \item Identify a heuristic notion of uncertainty using the pre-trained model.
  \item Define a score function $s(x, y) \in \mathbb{R}$. (Larger scores encode worse agreement between x and y.)
  \item Compute $\hat{q}$ as the $\frac{[(n+1)(1-\alpha)]}{n}$ quantile of the calibration scores $s_{1} = s(X_{1}, Y_{1}), ..., s_{n} = s(X_{n} Y_{n})$.
  \item Use this quantile to form the prediction sets for new examples:
  \begin{equation}
      {\Tau}(X_{test}) = \{y:s(X_{test},y) \leq \hat{q}\}
  \end{equation}
  where ${\Tau}(X_{test})$ represents a confidence set such that
  \begin{equation}
    1 - \alpha \leq \mathbb{P}(Y_{test} \in \Tau(X_{test})) \leq 1 - \alpha + \frac{1}{n+1}
  \end{equation}
  for some confidence level $\alpha$ and calibration set of unseen pairs of values and classes $(X_{1},Y_{1}), ... , (X_{n}, Y_{n})$  \cite{gentle_conf}
\end{enumerate} 

While this is the logic for discrete variables, this prediction method can also be applied in the case of regression. For example, take a regression model, $\hat{f}$, that predicts $y$ from $x$ and set our score function to the absolute error, such that $s(x,y) = |y - \hat{f}(x)|$. By using our calibration set as defined above we return $n$ error values. Then by taking the $1-\alpha$ quantile of our calibration scores, $\hat{q} = Quantile(s_{1}, s_{2}, ... , s_{n}; 1 - \alpha) = s_{q}$, we can create a confidence interval for the prediction: $\Tau(x) = [\hat{f}(x) - \hat{q}, \hat{f}(x) + \hat{q}]$. 

\begin{figure}[ht]
  \centering
  \includegraphics[width=0.8\linewidth]{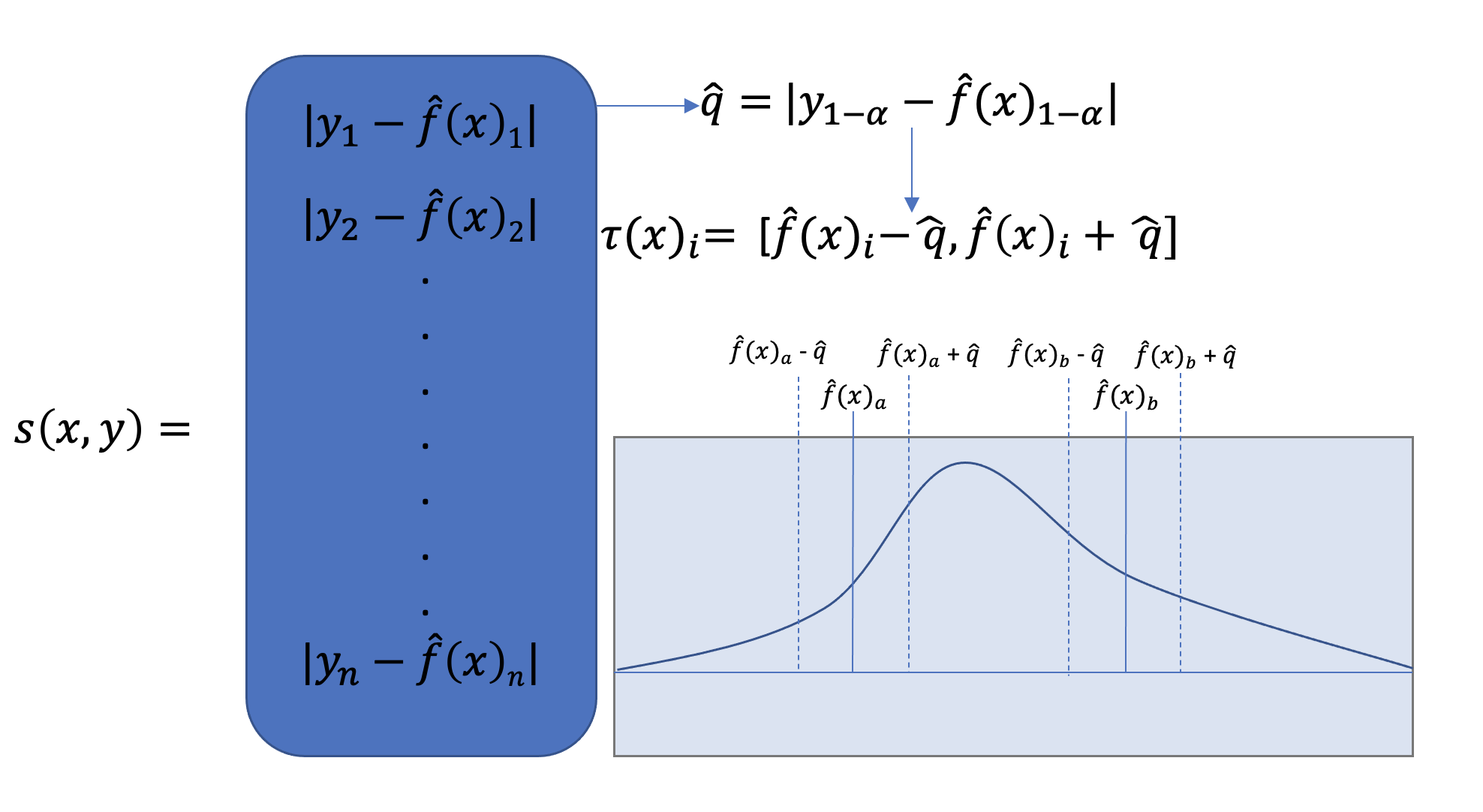}
  \caption{Example base confidence interval creation process for the Inductive Conformal method with constant interval size.}
  \label{fig:IC_pred}
\end{figure}

While this accomplishes the job of creating a confidence interval for each prediction, we can see that this prediction interval does not change based on $x$ and therefore does not give us a point-specific measure of confidence as shown by the illustration in Figure \ref{fig:IC_pred}. To get a dynamic confidence interval, we can go a step further. Instead of using the quantile of the absolute error, we can train a model, $\hat{r}$, that predicts $|y - \hat{f}(x)|$. If $\hat{r}$ was ideal we could just use $\hat{r}$ instead of $\hat{q}$ to construct a confidence interval, however in most cases the error model $\hat{r}$ has poor accuracy. To create a reliable estimate we can define a new score function: 
\begin{equation} 
    s(x,y) = \frac{|y - \hat{f}(x)|}{\hat{r}(x)}
\end{equation}
This new score function now outputs a new proportion that indicates error prediction inaccuracy and can be used as a corrective measure for the error. Rearranging equation (5) above we can see that $|y - \hat{f}(x)| = \hat{r}(x)s(x,y)$. Now we can use the same $1 - \alpha$ quantile as before to create a new $\hat{q}$ such that $\mathbb{P}[|Y_{test} - \hat{f}(X_{test})| \leq \hat{r}(X_{test})\hat{q}] \geq 1 - \alpha$. Using this we can create our final prediction:

\begin{equation}
    \Tau(X_{test}) = [\hat{f}(X_{test}) - \hat{r}(X_{test})\hat{q}, \hat{f}(X_{test}) + \hat{r}(X_{test})\hat{q}]
\end{equation}

In our experiment, as before we utilized an unchanged MEGNet Model which we trained on a subset of the data. We then compared the results from our initial trained model to an unused validation subset, and used the absolute error to get a new set of training data as the above method. We chose to use 95th percentile, or $\alpha = 0.05$, as our cut-off for $\hat{q}$.  Assuming a Gaussian distribution of the errors, this conformity factor allows us to have an estimated error that is approximately similar to our other error estimates which use 2 standard deviations (i.e. $\mathbb{P}[r < \hat{r}] \leq 1 - \alpha$).

\subsection{Deep Evidential Regression}

While the previous methods utilize extra models in order to estimate the prediction uncertainty, it would be ideal to train a single model that can estimate the uncertainty of its own predictions. A couple ways of accomplishing this feat include Bayesian Neural Networks, which will be discussed in the following section, or training a modified network that outputs the mean, $\mu(x)$, and variance, $\sigma^{2}(x)$, for a predicted value rather than an exact value. One exciting new application of the latter method comes in the form of Evidential Deep Learning \cite{amini2020deep, amini_2021}. Evidential Deep Learning builds off the concept of probabilistic learning. In probabilistic learning, for some data, $x$, and target, $y$, a model attempts to learn a mapping that creates a prediction $\mathbb{E}[y|x]$. This $\mathbb{E}[y]$ provides a point estimate of $y$ and we can further expand this to predict $Var[y]$ in order to estimate the uncertainty for each prediction. In Evidential Regression we assume that our target values, $y$, are drawn from some Gaussian distribution with some mean, $\mu$, and standard deviation, $\sigma^{2}$: 
\begin{equation}
y \sim Normal(\mu, \sigma^{2})
\end{equation}
These values will then allow the model to find a point estimate of $y$ along with an uncertainty estimate as for a Gaussian distribution $Var[y] = \sigma^{2}; \mathbb{E}[y] = \mu $. This approach is then taken a step further in Evidential Learning and the distribution parameters, $\mu$ and $\sigma^{2}$, are also probabilistically estimated. This is done by placing priors over each distribution parameter:
\begin{equation}
    \mu \sim Normal(\gamma, \sigma^{2}\nu^{-1})
\end{equation}

\begin{equation}
    \sigma^{2} \sim \Gamma^{-1}(\alpha, \beta)
\end{equation}

Rather than attempting to output the distribution parameters, $\mu$ and $\sigma^{2}$, the Evidential model instead outputs the parameters $\gamma, \nu, \alpha, \beta$, which allows $\mu$ and $\sigma^{2}$ to be drawn from the joint of the priors in Eq. (7) and Eq. (8):
$$ \mu,\sigma^{2} \sim NormalInvGamma(\gamma, \nu, \alpha, \beta)$$
This higher order distribution is what is called the Evidential Distribution, or Evidential Prior. Thus the evidential model is trained to output the parameters $\gamma, \nu, \alpha, \beta$ which define the evidential distribution. When sampled, the evidential distribution returns individual realizations of $\mu$ and $\sigma^{2}$, which are their own individual distributions defining the distribution from Eq. (6). The higher order distribution is called evidential since it has greater density in areas where there is more evidence in support of a given likelihood distribution observation. While many distributions can be picked over the likelihood, Amini et al.'s work \cite{amini2020deep} utilizes the Normal Inverse Gamma distribution due to it being a conjugate prior, which allows analytically computing the loss to be tractable. To give a further explanation, we can define $p(\mu,\sigma^{2}|\gamma, \nu, \alpha, \beta) = p(\theta|m)$ and $p(y|\mu,\sigma^{2}) = p(y|\theta)$. Then we can define:
\begin{equation}
p(\theta|y) = \frac{p(y|\theta)p(\theta)}{\int_{\theta'} p(y|\theta')p(\theta') d\theta'}
\end{equation}
Since our evidential prior, $p(\theta)$, is the same of the same family as our likelihood, $p(y)$, we can compute the integral in the denominator of Eq. (9) as part of the loss during training. In this experiment the final layer of the MEGNet Graph Neural Network was modified to output these Normal Inverse Gamma hyperparameters, giving 4 outputs for every target task. Thus the prediction and uncertainty were formulated according to the distribution moments:

\begin{equation}
Prediction: 
\mathbb{E}[\mu]= \gamma
\end{equation}

\begin{equation}
Uncertainty: 
Var[\mu]= \frac{\beta}{\nu(\alpha - 1)}
\end{equation}

The model was then trained using a dual-objective loss:
\begin{equation}
L(x) = L^{NLL}(x) + \lambda L^{R}(x)
\end{equation}
In the loss function defined by Eq. (12) $L^{NLL}(x)$ is the negative logarithm of model evidence while $L^{R}(x)$ is an evidence regularizer. In Amini et al.'s work these losses for the $i$-th prediction are defined as \cite{amini2020deep}:
\begin{equation}
L^{NLL}_{i} (x) = \frac{1}{2}\log(\frac{\pi}{\nu})-\alpha \log(2\beta(1+\nu)) + (\alpha + \frac{1}{2})\log((y_{i} - \gamma)^{2}\nu + 2\beta(1+\nu)) + \log(\frac{\Gamma(\alpha)}{\Gamma(\alpha + \frac{1}{2})})
\end{equation}
\begin{equation}
L^{R}_{i} (x) = |y_{i} - \mathbb{E}[\mu_{i}]| \cdot  \phi = |y_{i} - \gamma| \cdot (2\nu + \alpha) 
\end{equation}

This loss scales with with the total evidence of the inferred posterior ($\phi = 2\nu + \alpha$) and imposes a penalty whenever there is an error in the prediction. The $\lambda$ in the evidential loss function from Eq. (12) is known as the regularization coefficient and is used to trade off uncertainty inflation with model fit. If this regularization coefficient is set too low (e.g. $\lambda$ = 0) then the loss yields an over-confident estimate, while setting $\lambda$ too high results in over inflation. 


\begin{figure}[ht]
  \centering
  \includegraphics[width=0.9\linewidth]{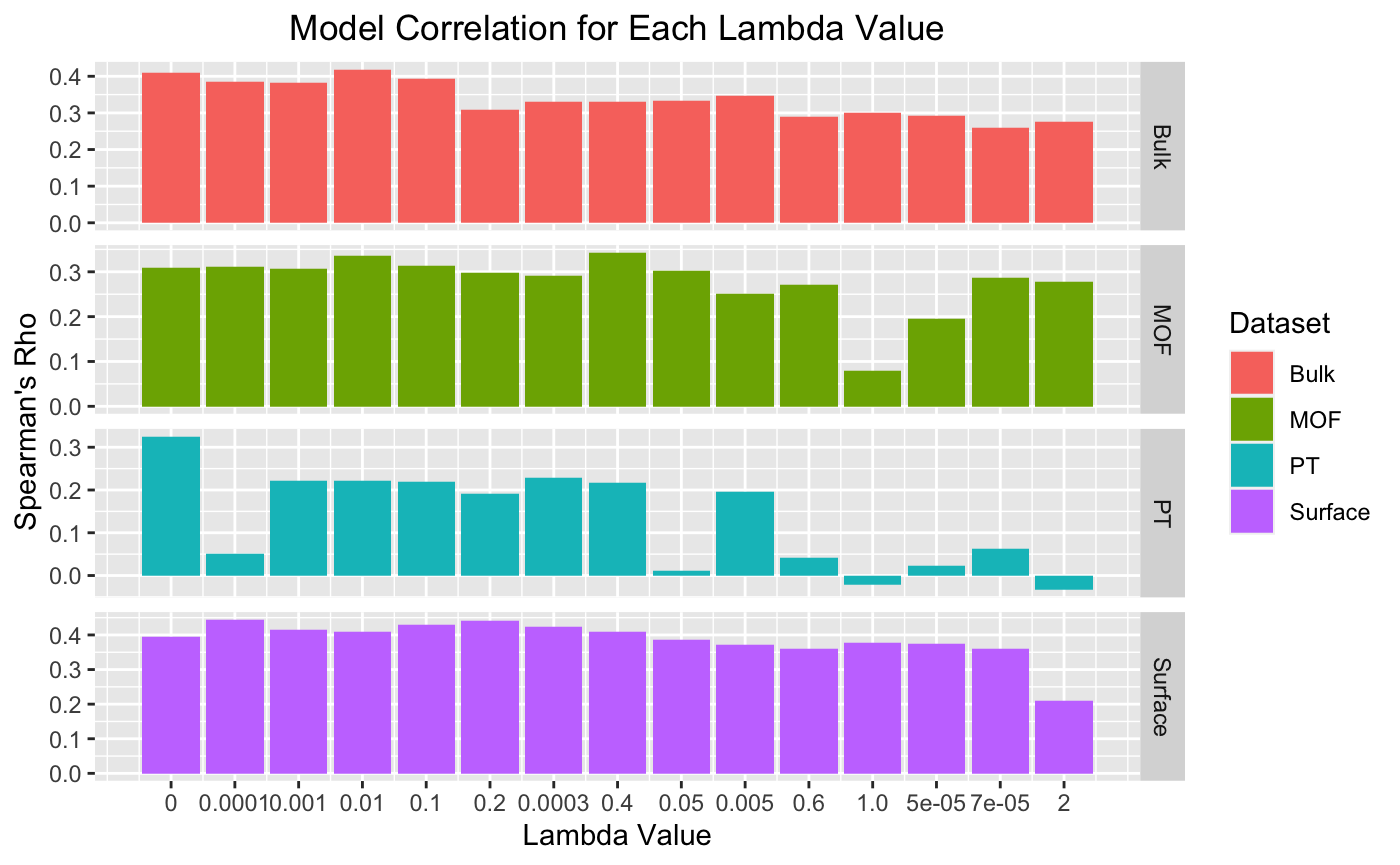}
  \caption{A higher value for Spearman's Rank Correlation suggests that there is a higher correlation between larger error values and larger uncertainty measures. By utilizing this metric to select between models we can choose the model where the variation in uncertainty most closely matches the variation in model error. The data sets from top to bottom are: (a) 3D bulk crystal structure formation energies, (b) 3D porous material MOF band gaps, (c) 0D sub-nanometer Pt cluster total energies, and (d) metal alloy surface absorption energies}
  \label{fig:Spearman's Rank Correlation Coefficient Values for each Value of Lambda and Data Set.}
\end{figure}

In our experiment we ran 15 trials using different regularization values for $\lambda$ with $\lambda \in [0, 0.00005, 0.00007, 0.0001, 0.0003, 0.001, 0.005, 0.01, 0.05, 0.1, 0.2, 0.4, 0.6, 1, 2]$. For each data set, we selected the model with the lambda value which produced the best Spearman's Rank Correlation to use as the comparison model. In the case of MOF data, we found that certain values of lambda produced a constant standard deviation, and as such no Spearman's correlation values were calculated for these values of lambda. For the Pt cluster, Bulk, MOF, and Surface data sets we found that the lambda values which produced the best results were 0, 0.0001, 0.01, and 0.00005, respectively (Figure 2).

\subsection{Delta Metric}
The last method of uncertainty generation for molecular discovery we will discuss is uncertainty through the use of a novel UQ measure developed by Karolev et al. \cite{delta_metric} called the $\Delta$ metric. This UQ measure is inspired by the K-nearest neighbors algorithm and adapted to applicability domain estimation in chemoinformatics. The $\Delta$ metric is a distance based measure that utilizes the average distance to the k closest training set points and compares it to a pre-defined threshold. Following the weighted k-nearest neighbor algorithm formulation, the delta metric uses the following formula for the $i$-th structure in the test set:

\begin{equation}
\Delta_{i} = \frac{\sum_{i}K_{ij}|\epsilon_j|}{\sum_{j}K_{ij}}
\end{equation}

Here $\epsilon_{j}$ represents the error between the target and prediction for the $j$th structure in the training set with $K_{ij}$ being the corresponding weight coefficient. This weight coefficient, $K_{ij}$, is a similarity measure between the $i$th and $j$th structures. This similarity measure is found using a kernel function proposed by Bartok et al. \cite{bartok2013representing} in the form of a normalized dot product of a global descriptor, $p$, to the $\zeta$ power:

\begin{equation}
    K_{ij} = (\frac{p_ip_j}{p_ip_j})^{\zeta}
\end{equation}

To featurize the structures for $p$, Karolev et al. utilized a smooth overlap of the atomic postion (SOAP) descriptor, as given by:

\begin{equation}
    p_{n_1n_2l} = \frac{\pi}{N^2}\sum_m\sum_{ij}(c^{i}_{n_1lm})^{\dagger}c^{j}_{n_2lm'}
\end{equation}

Here the variables, $c^{j}_{n_2lm'}$ and $c^{i}_{n_1lm}$, are the expansion coefficients in terms of the radial basis functions, $n_1$ and $n_2$, and the angular momentum channels, $l$, for the $i$-th and $j$-th atom, where N is the total number of atoms. According to the formulation given, the calculation for $\Delta$-metric only requires that the atomic structure(s) of interest, training set atomic structures, and the absolute prediction errors be provided. Many UQ methods require the use of multiple models which greatly increases the time and resources consumed for each new model employed, especially in the case of ensemble UQ method, thus being sub-optimal options when resources are limited. By only performing relatively simple calculations on ML model output, this metric receives an incredible boost in efficiency as compared to most the previous metrics described in this study, allowing it to be used for a single model output or in conjunction with an ensemble method.

\subsection{Evaluation criteria}

\paragraph{Root Mean Square Error}
One basic metric we can use to explore the effectiveness of different UQ methods is the Root Mean Square Error (RMSE). For a collection of materials, $D$, where the magnitude is $n = |D|$, the target value and its corresponding model estimate is represented by $y$ and $\hat{y}$ respectively, the RMSE formula is defined by:
\begin{equation}
RMSE(D) = \sqrt{\frac{\sum_{y \in D}(y - \hat{y})^{2} }{n}}
\end{equation}
We can utilize this metric to gauge whether the UQ correctly identifies the lowest error predictions by comparing the RMSE at different levels on uncertainty. We can split the error output for each model and data set into groups by the percentile rank of uncertainty (top 5, 10, 25, 50, 100\% of uncertainty values), and calculate the RMSE of each group. Ideally we should see a sharp monotonic decrease as the quantile grouping decreases, since we would expect the lowest 5\% of errors to in turn have the smallest uncertainty estimates followed by the 10\% and so on. While not a objective comparison metric like the following metrics, the RMSE groupings allows to visually observe if errors are following expected behaviour but along with how this behaviour differs between data sets.

\paragraph{Spearman's Rank Correlation Coefficient between Prediction Errors and UQ scores as the metric for evaluating UQ methods}
Given an uncertainty estimator, it is expected that for predictions with low uncertainty there will be a corresponding relatively low true error. Essentially, given some model $M$ with two inputs $a$ and $b$ for which the corresponding uncertainty estimates $U(a)$ and $U(b)$ follow $ U(a) < U(b)$, then we would expect the prediction $M(a)$ to be more accurate than $M(b)$. The extent to which a model follows this rule can be measured utilizing Spearman's Rank Correlation Coefficient. While Pearson's Correlation Coefficient is a very common measure of correlation, Spearman's was favorable in this case as it measures the monotonic relationship between two variables and the relationship between the accuracy of a prediction and it's corresponding uncertainty is not necessarily linear. Unlike Pearson's Correlation which uses the values themselves, Spearman's Correlation uses the ranks of the values. Given two vectors $X$ and $Y$, we can create ranked vectors $x$ and $y$ which assign an integer rank to each value in their corresponding vector in ascending order. The correlation coefficient $\rho$ is then defined by the function where $\sigma_{x}$, $\sigma_{y}$, $\sigma_{xy}$ define the standard deviation of $x$, the standard deviation of $y$, and the covariance between them:
\begin{equation}
\rho(X,Y) = \frac{\sigma_{xy}}{\sigma_{x}\sigma_{y}}
\end{equation}
$\rho(X,Y)$ can be calculated using the following formulas in the case where the data does not include ties and in the case where there are ranked ties, respectively. $d_{i}$ is calculated as the difference in paired ranks, with $i$ representing a paired rank, and $n$ is the total length of the vectors:
\begin{equation}
\rho(X,Y) = 1 - \frac{6\sum d_{i}^{2}}{n(n^{2}-1)}
\end{equation}

\begin{equation}
\rho(X,Y) = \frac{\sum_{i} (x_{i} - \bar{x})(y_{i} - \bar{y}) }{\sqrt{\sum_{i} (x_{i} - \bar{x})^{2}(y_{i} - \bar{y})^{2} }}
\end{equation}
In the case where $X$ and $Y$ have the exact same rankings $\rho(X,Y) = 1$ and in the case they have opposite rankings $\rho(X,Y) = -1$. Since we believe the errors will be approximated normally distributed we do not expect a rank correlation coefficient of -1 or 1. In our evaluation, the rank correlation coefficient is calculated by comparing a vector of absolute errors and a vector of uncertainty measures. Since this metric compares the ranks of values and not the values themselves, it allows us to compare methods regardless of whether they produce a true standard deviation or not.

\paragraph{Miscalibration Area}
Another way to evaluate these uncertainty estimate methods is to check whether the predicted uncertainties are similar in magnitude to the true errors observed. This is referred to as calibration. Tran et al. \cite{tran2020methods} proposed comparing the fraction of observed errors that fall within n standard deviations of the mean to what is expected for a Gaussian random variable with variance equal to the uncertainty prediction of the uncertainty estimation methods.
Given a perfect uncertainty estimator, then the theoretical fraction, or the theoretical cumulative distribution (CD) of points, should exactly match the observed fraction or CD for any n. The miscalibration area itself is found to be the area between the true curve of observed versus expected fractions and the parity line. As such a perfect uncertainty estimation method will have an exact one to one relationship with the expected expected fraction, giving a miscalibration area of 0. This metric provides a measure of whether the method in question is systematically over or under-confident when estimating the error, and as such it is possible to get a score of 0 if approximately half of the data points are under-confident and the other half are over-confident.

\paragraph{Negative Log Likelihood}
The next metric we utilize to evaluate the uncertainty estimation performance is the Negative Log Likelihood (NLL). Here we are measuring the NLL of the observed model errors under the assumption that they are normally distributed with a mean of 0, with the variances given by the uncertainty estimate represented by $U(x)$. Given a collection of materials, $D$, where the magnitude is $n = |D|$, then the average NLL can be defined as: \cite{amini_2021}
\begin{equation}
NLL(D) = \frac{1}{2n}\sum_{x,y \in D}ln(2\pi) + ln(U(x)) + \frac{(M(x) - y)^{2}}{U(x)}
\end{equation}
This NLL value is averaged using, $n$, as it allows comparison between data sets by accounting for bias when dealing with larger data sets.

\paragraph{Calibrated Negative Log Likelihood}
The challenge with using NLL as described above is that for uncertainty estimation methods that are not intended to be used as estimations of variance (like the measure in the Inductive Conformal method), the NLL will not show any useful results. One solution to apply NLL to other methods, as described in Hirschfeld et al.\cite{janet2019quantitative, hirschfeld2020uncertainty}, is to perform a transformation on the uncertainty estimates so they resemble variances. Instead of taking $\hat{\sigma}^2(x) := U(x)$, this calibration is accomplished by assuming that $\hat{\sigma}^2$ and $U(x)$ are linearly related:
\begin{equation}
\hat{\sigma}^2(x) := aU(x) + b
\end{equation}

So to compute the the Calibrated NLL (cNLL) for each data set and method, a set of scalars, $a_*$ and $b_*$, are found which minimize the NLL of errors which minimize the set of errors in each respective validation set, $D_{val}$.
\begin{equation}
a_*,b_* = argmin_{a,b}\frac{1}{2}\sum_{x,y \in D_{val}}ln(2\pi) + ln(aU(x) + b) + \frac{(M(x) - y)^{2}}{aU(x) + b}
\end{equation}

Then using these scalars we can compute the cNLL for the test set, $D_{test}$ in a similar manner as above:
\begin{equation}
cNLL(D_{test}) = \frac{1}{2n}\sum_{x,y \in D_{test} }ln(2\pi) + ln(a_*U(x) + b_*) + \frac{(M(x) - y)^{2}}{a_*U(x) + b_*}
\end{equation}

\subsection{Data sets}

The data sets we used were the same as those used by Fung et al.~\cite{matDL_pap} to benchmark graph neural networks for materials chemistry. These data sets were chosen by the researchers to reasonably represent a variety of different types of inorganic materials ranging from 0-3D and provided a robust data set for the benchmarking or performance of different materials science neural networks, and as such was seen to be great set of data for benchmarking uncertainty estimation. The information about the four datasets is presented in Table~\ref{table:dataset}. The readers are requested to refer to Fung et al.~\cite{matDL_pap} for more information on each data set. For each data set and model a 75\%/10\%/15\% train/validation/test split was utilized in training and comparison.

\begin{table}[t]
\caption{Details of the four benchmark datasets used in this work.}
\begin{center}
\begin{tabular}{l c c c c}
\hline
\textbf{Dataset } & \textbf{Unit} & 
\textbf{Source} & 
\textbf{\# of elements} &
\multicolumn{1}{c}{\textbf{\# of samples}} \\ \hline

Bulk Materials Formation Energy & eV/atom             & MaterialsProject \cite{jain2013commentary}                 & 87        & 36839              \\ 
Alloy Surface Adsorption Energy & eV         & CatHub \cite{mamun2019high}    & 42               & 37334           \\ 
Pt-cluster Total Energy & eV         & Literature \cite{fung2017exploring}    & 1               & 19801           \\ 
MOF Band Gap  & eV        & QMOF \cite{rosen2021machine}                & 78        & 18321        \\ \hline

\end{tabular}
\label{table:dataset}
\end{center}
\end{table}

\section{Results}


The following section will showcase and explain the UQ performances of different algorithms using the metrics mentioned above. We utilize RMSE as an exploratory metric to observe differences between models and data sets, while all the following four metrics provide objective values to compare the different approaches for UQ of materials property prediction models.

\subsection{Spearman's Rank Correlation}
\begin{figure}[ht]
  \centering
  \includegraphics[width=0.8\linewidth]{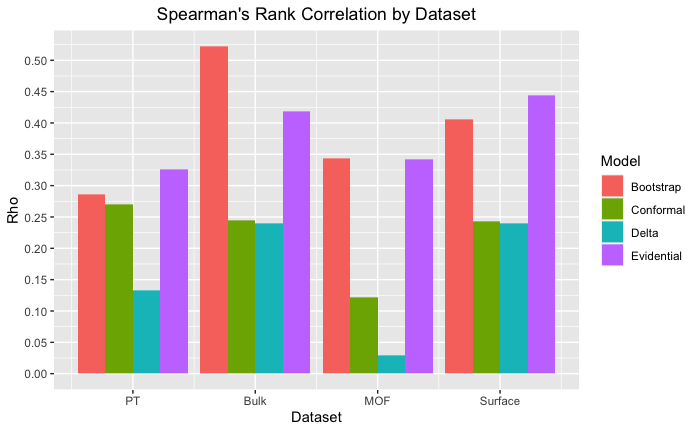}
  \caption{Spearman's Rank Correlation Coefficient (Rho) calculated for all data sets and UQ methods evaluated in this study. The correlation was calculated between the absolute error and uncertainty produced by each model. The data sets from left to right: (a) 0D sub-nanometer Pt cluster total energies, (b) 3D bulk crystal structure formation energies, (c) 3D porous material MOF band gaps, and (d) metal alloy surface absorption energies}
  \label{fig:Spearman's Ranked Correlation Coefficient for each Model and Dataset.}
\end{figure}

Figure \ref{fig:Spearman's Ranked Correlation Coefficient for each Model and Dataset.} shows Spearman's Rank Correlation Coefficient for each method and data set. From the figure, we can immediately see variation in the correlation based on both the method type and the data set. The Ensemble method had the highest variation between data sets with an approximately 82\% increase in correlation from the Pt Cluster data set to the Bulk Crystal Formation energy data set. The Conformal method, on the other hand, maintained a relatively stable correlation, with only the MOF data set causing a large change. While the MOF data set for the Conformal method experienced a relatively large drop in correlation ($\rho \approx 0.122$) compared to the rest, there was only an approximate difference of 0.027 between its largest correlation coefficient with the Pt data set ($\rho \approx 0.269$) and its second smallest coefficient with the Surface data set ($\rho \approx 0.243$). The variation between correlation across different data sets in the Evidential method lies somewhere between the two, with similar correlations between the Bulk and Surface data set and a similar drop in correlation with the Pt and MOF data sets compared to the previous two. Interestingly, the Pt data saw strong drops in correlation for the Evidential and the Ensemble methods, however, the largest correlation was found for the Evidential method. Through just the rank correlation is difficult to conclusively state which model performed at the greatest level though the Evidential model performed strongly as it saw the top correlation with the Pt data a slightly higher correlation with the Surface data, while the Ensemble model had the best performance with the Bulk data. The MOF data saw an extremely slight improvement with the Evidential method ($\rho \approx 0.343$) over the Ensemble method ($\rho \approx 0.342$), but with a difference of approximately 0.0005 the methods were essentially tied in this instance.  As we can see in Figure \ref{fig:Spearman's Ranked Correlation Coefficient for each Model and Dataset.} the Conformal method produced the lowest ranked correlation in each data set, failing to beat any models, likely stemming from the use of calculated distances to create uncertainty intervals rather than any true variance measure.

\subsection{Root Mean Square Error}
\begin{figure}[ht]
  \centering
  \includegraphics[width=0.85\linewidth]{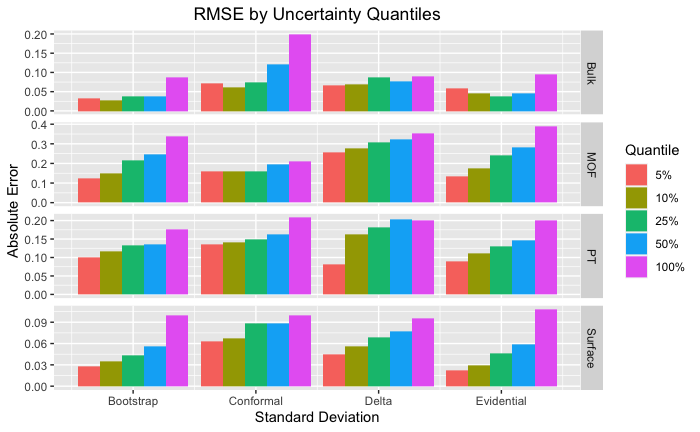}
  \caption{RMSE calculated for all data sets and UQ methods evaluated in this study. The RMSE was calculated for the predictions with the 100, 50, 25, 10, and 5\% lowest uncertainty generated by the UQ method. The data sets from top to bottom are: (a) 3D bulk crystal structure formation energies, (b) 3D porous material MOF band gaps, (c) 0D sub-nanometer Pt cluster total energies, and (d) metal alloy surface absorption energies}
  \label{fig:RMSE by Quantile for Each Model and Dataset.}
\end{figure}

As stated above, a sharp monotonic decrease in the RMSE between the different quantile uncertainty groups suggests that the UQ method correctly identifies the predictions in which it has the lowest confidence. Figure \ref{fig:RMSE by Quantile for Each Model and Dataset.} illustrates the difference in RMSE when calculated on different subsets of the test data which the UQ method is most uncertain about. We can see that the results for each model can vary strongly between different data sets, showing the importance of using multiple data sets when comparing UQ methods, as a single data set may lead to improper conclusions about the methods. For instance, looking at the MOF data set, even though the Evidential, Delta, and Ensemble UQ methods have the higher total RMSE error, they seem to best detect which predictions are the most uncertain as all three graphs have strong decreases in RMSE as the quantile grouping decreases. Though it had the lowest total RMSE showing that the predictions were more accurate than the former methods, the Inductive Conformal method ended up with an almost constant RMSE between all quantile groups. On the other hand, the Inductive Conformal method seemed to have the best trend for the Bulk data set with distinct `steps' being visible as the groupings decreased, while the other three UQ methods had almost parabolic trends in RMSE between groups. With the MOF, Pt, and Surface data sets both the Ensemble and the Evidential methods had relatively similar decreases in RMSE for each grouping, showcasing the desired monotonic decreases between uncertainty quantile groupings. The Delta metric also showed a decreasing trend in RMSE for each quantile "step" with the MOF and Surface data sets, however it did not perform as well as the two aforementioned method with the PT data set and produced higher RMSEs for a majority of the quantile groups. The Evidential method had the an overall greater RMSE with the MOF, Pt, and Surface data sets than the Ensemble method but the smallest RMSE for the 5\% quantile with both Surface and Pt data sets. Across all data sets except for the Bulk data, the Inductive Conformal method seems to have struggled the most as it failed to have a sharp decrease between different uncertainty groupings with only slight decreases in groupings with the Pt and Surface data sets.

\subsection{Miscalibration Area}
\begin{figure}[ht]
  \centering
  \includegraphics[width=0.8\linewidth]{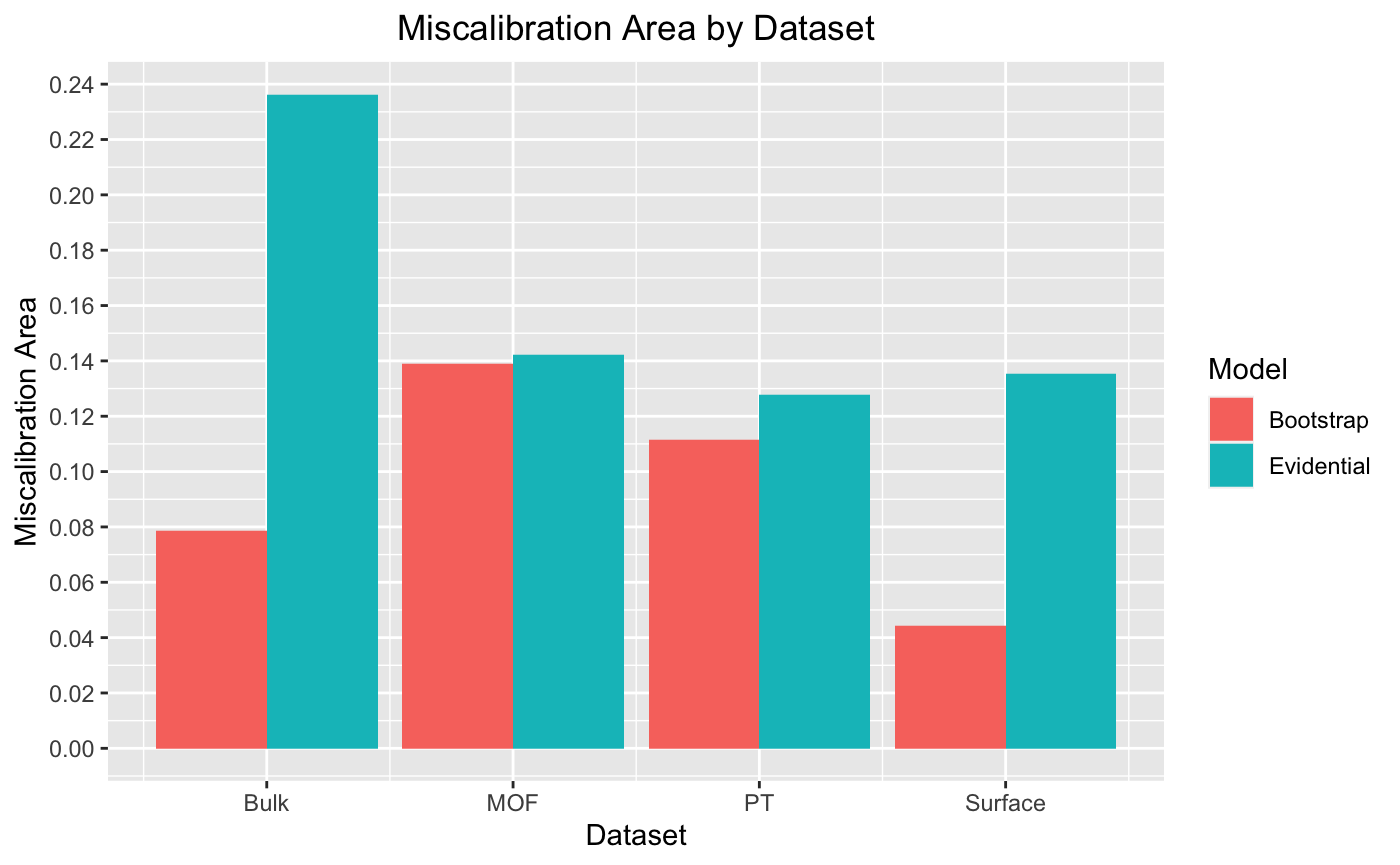}
  \caption{Miscalibration area calculated for all data sets and variance-based UQ methods evaluated in this study. The miscalibration area for the Conformal method was not shown as the uncertainty measures outputted are not meant to be utilized as true variances. The data sets from left to right: (a) 3D bulk crystal structure formation energies, (b) 3D porous material MOF band gaps, (c) 0D sub-nanometer Pt cluster total energies, and (d) metal alloy surface absorption energies}
  \label{fig:Miscalibration Areas for Each Model and Dataset.}
\end{figure}

Miscalibration area, allows us to analyze which methods systematically overestimate or underestimate the uncertainty of predictions. Here, a score of 0 indicates a perfect alignment between the expected theoretical Gaussian cumulative distribution and the observed distribution of errors while a score of 0.5 indicates maximal misalignment. 
It is important to keep in mind that it is theoretically still possible to get a score of 0 if a method equally overestimates and underestimates prediction uncertainties. We can see from Figure \ref{fig:Miscalibration Areas for Each Model and Dataset.} that for each of the data sets the Ensemble method produces the lowest miscalibration area. The MOF data set provided the smallest variation between the Evidential ($miscal_area \approx 0.142$) and Ensemble ($miscal_area \approx 0.139$) methods with an approximate difference of 0.003, while the Bulk data set produced the highest variation between the Evidential ($miscal_area \approx 0.236$) and Ensemble ($miscal_area \approx 0.078$) methods with an approximate difference of 0.158 between the two. Note that the miscalibration areas for the Conformal or Delta methods are not shown as the uncertainties here are not meant to be used as variances and would produce incoherent results. We can see that this difference is greatest in the Bulk data set which is sensible as the Evidential model had the worst performance out of three methods in RMSE groupings (\ref{fig:RMSE by Quantile for Each Model and Dataset.}) for the Bulk data set.

It seems that the Evidential model tends to over/under estimate the uncertainty of predictions to a greater degree than the Ensemble model, though the degree to which it does varies by data set. It is interesting to note that in the MOF, Pt, and Surface data sets the miscalibration areas were relatively constant with a < 0.015 difference between the the largest and smallest miscalibration areas of the three sets, but there is a drastic increase in miscalibration area with the Bulk data set compared to the rest with a difference of approximately 0.094 between the Bulk miscalibration area and the next largest area with the MOF data. The Ensemble method seemed to vary more across data sets with an approximate difference of 0.094 between its largest value in the MOF data set and its smallest in the Surface data set and an approximately 0.067 difference in miscalibration area from the method's second largest area with the Pt data set and the area from the Surface data set.

\FloatBarrier

\subsection{Negative Log Likelihood}
\begin{figure}[ht]
  \centering
  \includegraphics[width=0.8\linewidth]{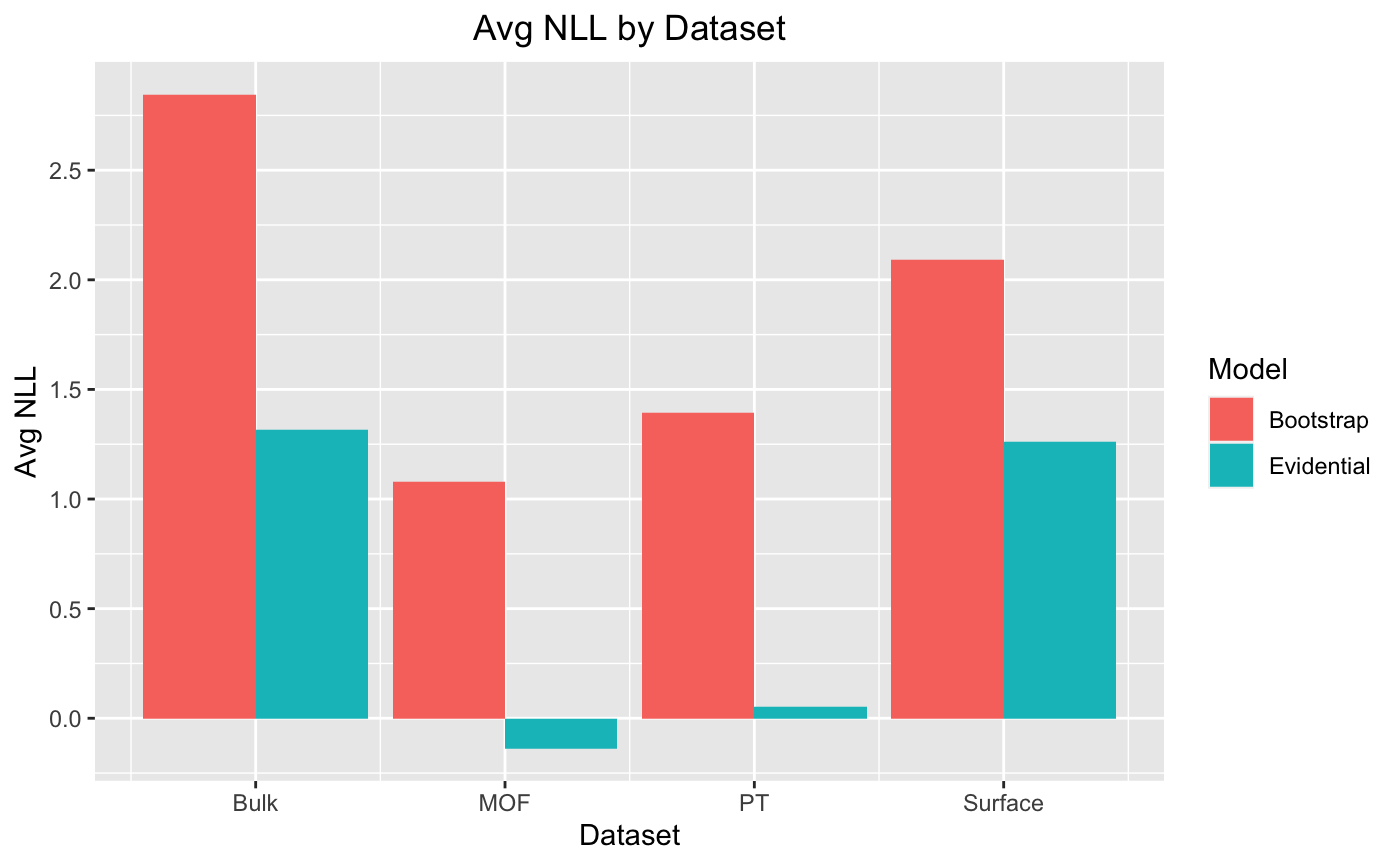}
  \caption{Gaussian Negative Log Likelihood calculated for all data sets and variance based UQ methods evaluated in this study. The negative log likelihood for the Conformal method was not shown as the uncertainty measures outputted are not meant to be utilized as true variances. The data sets from left to right: (a) 3D bulk crystal structure formation energies, (b) 3D porous material MOF band gaps, (c) 0D sub-nanometer Pt cluster total energies, and (d) metal alloy surface absorption energies}
  \label{fig:Negative Log Likelihood Values for Each Model and Dataset.}
\end{figure}

Figure \ref{fig:Negative Log Likelihood Values for Each Model and Dataset.} displays the results for the uncalibrated Negative Log Likelihood. As with the Miscalibration Area, the NLLs for the Conformal and Delta methods are not displayed since the uncertainty measure for those methods is not meant to be an estimate of variance and the NLL calculation works under the assumption that the normally distributed errors have variance equal to the method's predicted uncertainty measure. It is also important to note that the accuracy of the method can influence the NLL as a more accurate method with lower errors will have a higher maximum likelihood and therefore a lower NLL. 

From Figure \ref{fig:Negative Log Likelihood Values for Each Model and Dataset.} we can see that the Evidential method performs significantly better than the Ensemble method for each data set. The fact that the Evidential method was superior to the Ensemble method for every data set was quite surprising, especially considering that in each data set the total RMSE was higher for the Evidential model than the Ensemble model as seen in Figure \ref{fig:RMSE by Quantile for Each Model and Dataset.}, limiting the possible minimum value for the NLL of the Evidential model more than that of the Ensemble model.The Bulk data set had the largest respective average NLL out of all data sets for both the Evidential and Ensemble method, which can be expected as both methods had their weakest performance in RMSE groupings. We can also see significant differences between in the NLL between half of the data sets for the Evidential method. For Evidential method the MOF and and Pt data sets having relatively similar NLL values and the Surface and Bulk achieving very similar between each other, mirroring the similarities across datasets in correlation from Figure \ref{fig:Spearman's Ranked Correlation Coefficient for each Model and Dataset.}.

\subsection{Calibrated Negative Log Likelihood}
\begin{figure}[ht]
  \centering
  \includegraphics[width=0.8\linewidth]{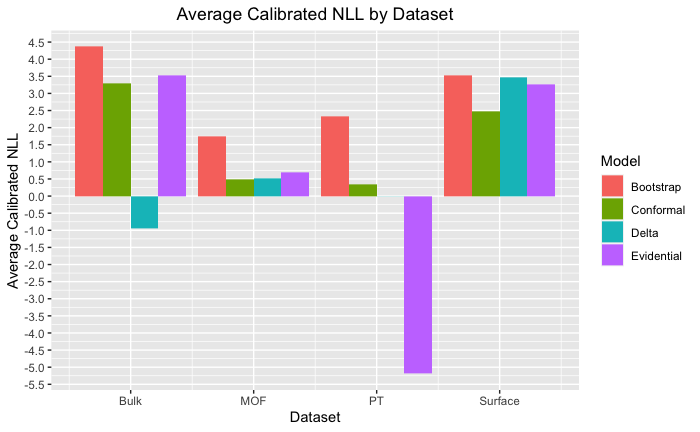}
  \caption{Gaussian Calibrated Negative Log Likelihood calculated for all data sets and variance-based UQ methods evaluated in this study. The data sets from left to right: (a) 3D bulk crystal structure formation energies, (b) 3D porous material MOF band gaps, (c) 0D sub-nanometer Pt cluster total energies, and (d) metal alloy surface absorption energies}
  \label{fig:Calibrated Negative Log Likelihood Values for Each Model and Dataset.}
\end{figure}

Calibrating the uncertainty measures so they more closely resemble true variances allows us to directly compare the Conformal method to the Evidential and the Ensemble UQ methods, as the uncertainty estimates produced by the Conformal method are not meant to be used as variances. This is a more forgiving evaluation than the previous NLL measure as it uses an optimized measure of $\hat{\sigma}^2(x)$ since $\hat{\sigma}^2(x) := aU(x) + b$ (Eq. 21) rather than directly utilizing  $U(x)$ as in NLL. Figure \ref{fig:Calibrated Negative Log Likelihood Values for Each Model and Dataset.} shows the calibrated NLL (cNLL) for each UQ method after the uncertainty measures were calibrated. The calibration seemed to have mixed results across models leading to improvements for some models and data sets and decreases in others. We can see this with respect to the Evidential method, where the cNLL for the Pt data set was greatly decreased from the original NLL value, however, the cNLL value for the Bulk, MOF, and the Surface data sets was increased. It seems that there was some variation between data sets on which model is performing the best, as in the Pt data set, the Evidential method significantly outperforms the other three UQ methods, however, with the rest it fails to outperform either Delta or Inductive Conformal method. For the MOF data set, the Inductive Conformal and Delta methods are extremely close at cNLL values of approximately 0.492 and 0.516 respectively. With the other two data sets, Bulk and Surface, the performance differences between the two methods varied significantly.  It can be seen that for the former data set the Delta metric outperforms the rest significantly, yet with the latter data set the Inductive Conformal method significantly outperformed all other methods with the Delta method barely coming in third place. It should be noted that calibration is sensitive to the choice of calibration data, the validation set for each data set and method in this case, and can harm the performance of previously well-calibrated methods. This effect can be seen when comparing the NLL to the cNLL in Figure \ref{fig:Calibrated Negative Log Likelihood Values for Each Model and Dataset.}d as the NLL value increased approximately 69.14\% and 157.53\% for the Ensemble and the Evidential methods, respectively.

\FloatBarrier

\section{Discussion and Conclusion}


Our study attempts to compare and analyze the performance of three different UQ methods for materials property prediction that can easily be applied to most NN approaches with minimal adjustment of the existing architecture. With the Inductive Conformal and Evidential methodologies being relatively new developments in UQ, we utilized the classic Bootstrap Ensemble method to benchmark their performance alongside a UQ staple. We used four different data sets to showcase the performance and stability across multiple tasks, an important factor when considering which method to use. The four main metrics used in our evaluation are: Spearman’s rank correlation coefficient, which measured how closely the ranking of uncertainty estimates matches the ranking of errors; the miscalibration area, which measures whether the uncertainties tend to over/underestimate the errors by modeling the errors as normal variables with variances equal to the uncertainty estimates; the average negative log likelihood of the observed value, using models’ predicted values and uncertainty estimates as means and variances of normal distributions; and a calibrated negative log likelihood, where the uncertainty estimates are calibrated using the validation set to create an optimized uncertainty measure that is linearly related to the variance. We additionally employed a graph of RMSEs grouped by different uncertainty quantiles to compare the accuracy of the models and how they varied across different tasks.

From Figure \ref{fig:RMSE by Quantile for Each Model and Dataset.} we can that the UQ model performance tended to vary across data sets with different degrees. With regards to the overall accuracy of the models, the total RMSE for the each data set except the Bulk data was slightly higher for the Evidential model compared to the Ensemble model, which is likely since the Evidential model requires a full change in the output layer and loss function of the MEGNet model. The Inductive Conformal method had a much higher RMSE than the other methods with the Bulk data, which could stem from the fact that it uses point estimates rather than mean estimates and had only a single prediction model leaving it vulnerable to the choice of validation and training data. Concerning the actual uncertainty, we can see significant differences in performance between data sets. It seems that all methods have the most trouble with the Bulk data set when creating uncertainty estimates as it generally has amongst the worst results for the miscalibration area, RMSE grouping trends, NLL, and cNLL out of the four data sets. On the other hand, for the Spearman's rank correlation the Bulk data set had a significantly higher average rho between the three methods than any of the other data sets with an approximately 40.51\% increase from the Pt data set to the Bulk data set. Though the Conformal method was the most consistent out of three methods, pit showed fairly weak performance with a correlation coefficient range of 0.1218 to 0.2695. The Delta method seems to have the weakest performance in correlation, performing on par with the Inductive Conformal in the Bulk and Surface data sets, yet being significantly outperformed in the PT and MOF data sets with correlations of approximately 0.13 and 0.29, respectively.

The performance of UQ methods is also inconsistent across different evaluation metrics. We can see from Figure \ref{fig:Miscalibration Areas for Each Model and Dataset.} that the Evidential method has the larger miscalibration area for each data set, however, from Figure \ref{fig:Negative Log Likelihood Values for Each Model and Dataset.} we can see that the Evidential method now significantly outperforms the Ensemble method in each data set. If we look at figure \ref{fig:Calibrated Negative Log Likelihood Values for Each Model and Dataset.} it seems that the Inductive Conformal method has a superior cNLL value for each data set except the Pt data and that the Delta method has superior performance with the Bulk and MOF data sets, however it can be that calibration actually harms performance for methods that are already well calibrated. This seems to be the case with the Evidential method as its NLL values actually increased for every data set except the Pt data and when we compare the true NLL for the Evidential Method to the cNLL for the Inductive Conformal method we can see that the Evidential method is actually also superior for the other three data sets. On the other hand, when we compare the Evidential NLL to the Delta cNLL, we can see that the Delta method outperforms the Evidential method in the Bulk data set regardless of calibration. The Delta method also had strong performance in the PT data set, only placing second to the Evidential cNLL and outperforming the Evidential NLL, a surprising result due the Delta methods weak results in Spearman's correlation tests. Overall, the Delta method showed relatively strong, consistent results across the the Bulk, MOF, and PT data sets with all three having a < 0.52 correlation coefficient value. For these three data sets the Delta metric ranked either first or second in cNLL out of all models, though it seemed to have issues with the errors in the Surface data set.It is interesting to note that the miscalibration are for Evidential method with the Bulk data set was significantly higher than any of the others measured, however the Evidential method still achieved a lower NLL than the Ensemble method with the aforementioned data set. This could be due to the Evidential method having a tendency to over/underestimate the error in one direction, while the Ensemble model was over/underestimating more equally throughout the data set.

Though the Inductive Conformal, Delta, and the Ensemble methods each had their benefits in performance, our benchmark study shows that the Evidential UQ method is a strong candidate for a first UQ approach in future UQ work, with its strong and consistent performance in the NLL and Spearman's Rank Correlation analyses. One important consideration of the Evidential model is the resources it uses. Unlike the Ensemble method where one has to train multiple models on different samples or the Conformal method which requires the training of two different models, the Evidential method requires the training of only a single model. The Delta metric also only requires the training of a single model and showed promise with relatively strong performance in cNLL, however the weak correlation coefficients and the varied RMSE results currently give the edge to the Evidential method. The Evidential method achieved relatively consistent performance across different data sets, especially with the MOF and PT data along with the Bulk and Surface data achieving very similar results to each other across most metrics, respectively, lessening the possible variation in performance when working with a novel set of data. Moreover, when looking at both cNLL and NLL which reflect both UQ and regression accuracy, the uncalibrated Evidential method mostly outperformed all three of of the other methods. The slightly larger RMSE groupings for the Evidential method for some of the data sets may be a slight drawback when compared to the other two methods, however, the strong performance in Spearman's rank correlation shows that the ranked uncertainty is still strongly correlated to the errors. Future work can expand on this study to benchmark UQ methods for classification tasks and benchmark utilizing several different models to see how the baseline model architecture caused variation in performance across methods. Moreover, the novel Delta metric achieved promising cNLL results and it would be interesting to determine how the delta method performance changes and if the relatively large variance in results is significantly decreased when combined with a small ensemble of models.

\section{Data Availability}

The data that supports the findings of this study is openly downloadable as stated in reference \cite{matDL_pap}.

\section{Contribution}
Conceptualization, J.H.; methodology,D.V. and J.H.; software, D.V.; resources, J.H.; writing--original draft preparation, D.V., J.H., R.D., S.O.; writing--review and editing,  J.H, D.V.; visualization, D.V., R.D.; supervision, J.H.

\section{Acknowledgement}
Research reported in this work was supported in part by NSF under grants 1940099 and 1905775. The views, perspective, and content do not necessarily represent the official views of the SC EPSCoR Program nor those of the NSF. 


\bibliographystyle{unsrt}  
\bibliography{references}

\begin{thebibliography}{10}

\bibitem{schleder2019dft}
Gabriel~R Schleder, Antonio~CM Padilha, Carlos~Mera Acosta, Marcio Costa, and
  Adalberto Fazzio.
\newblock From dft to machine learning: recent approaches to materials
  science--a review.
\newblock {\em Journal of Physics: Materials}, 2(3):032001, 2019.

\bibitem{korzdorfer2014organic}
Thomas K{\"o}rzd{\"o}rfer and Jean-Luc Br{\'e}das.
\newblock Organic electronic materials: recent advances in the dft description
  of the ground and excited states using tuned range-separated hybrid
  functionals.
\newblock {\em Accounts of chemical research}, 47(11):3284--3291, 2014.

\bibitem{maurer2019advances}
Reinhard~J Maurer, Christoph Freysoldt, Anthony~M Reilly, Jan~Gerit
  Brandenburg, Oliver~T Hofmann, Torbj{\"o}rn Bj{\"o}rkman, S{\'e}bastien
  Leb{\`e}gue, and Alexandre Tkatchenko.
\newblock Advances in density-functional calculations for materials modeling.
\newblock {\em Annual Review of Materials Research}, 49:1--30, 2019.

\bibitem{kohn1965self}
Walter Kohn and Lu~Jeu Sham.
\newblock Self-consistent equations including exchange and correlation effects.
\newblock {\em Physical review}, 140(4A):A1133, 1965.

\bibitem{hohenberg1964inhomogeneous}
Pierre Hohenberg and Walter Kohn.
\newblock Inhomogeneous electron gas.
\newblock {\em Physical review}, 136(3B):B864, 1964.

\bibitem{ding2009molecular}
Lifeng Ding, Ruslan~L Davidchack, and Jingzhe Pan.
\newblock A molecular dynamics study of sintering between nanoparticles.
\newblock {\em Computational Materials Science}, 45(2):247--256, 2009.

\bibitem{steinbach2009phase}
Ingo Steinbach.
\newblock Phase-field models in materials science.
\newblock {\em Modelling and simulation in materials science and engineering},
  17(7):073001, 2009.

\bibitem{fallah2012phase}
V~Fallah, M~Amoorezaei, N~Provatas, SF~Corbin, and A~Khajepour.
\newblock Phase-field simulation of solidification morphology in laser powder
  deposition of ti--nb alloys.
\newblock {\em Acta Materialia}, 60(4):1633--1646, 2012.

\bibitem{zhao2021screening}
Xiaojun Zhao, Peng Wang, Erfei Lv, Chongchong Wu, Kai Ma, Zhengyang Gao, Ian~D
  Gates, and Weijie Yang.
\newblock Screening mxenes for novel anode material of lithium-ion batteries
  with high capacity and stability: A dft calculation.
\newblock {\em Applied Surface Science}, 569:151050, 2021.

\bibitem{zhang2019high}
Xu~Zhang, An~Chen, and Zhen Zhou.
\newblock High-throughput computational screening of layered and
  two-dimensional materials.
\newblock {\em Wiley Interdisciplinary Reviews: Computational Molecular
  Science}, 9(1):e1385, 2019.

\bibitem{hill2016materials}
Joanne Hill, Gregory Mulholland, Kristin Persson, Ram Seshadri, Chris
  Wolverton, and Bryce Meredig.
\newblock Materials science with large-scale data and informatics: Unlocking
  new opportunities.
\newblock {\em Mrs Bulletin}, 41(5):399--409, 2016.

\bibitem{himanen2019data}
Lauri Himanen, Amber Geurts, Adam~Stuart Foster, and Patrick Rinke.
\newblock Data-driven materials science: status, challenges, and perspectives.
\newblock {\em Advanced Science}, 6(21):1900808, 2019.

\bibitem{agrawal2016perspective}
Ankit Agrawal and Alok Choudhary.
\newblock Perspective: Materials informatics and big data: Realization of the
  “fourth paradigm” of science in materials science.
\newblock {\em Apl Materials}, 4(5):053208, 2016.

\bibitem{calderon2015aflow}
Camilo~E Calderon, Jose~J Plata, Cormac Toher, Corey Oses, Ohad Levy, Marco
  Fornari, Amir Natan, Michael~J Mehl, Gus Hart, Marco~Buongiorno Nardelli,
  et~al.
\newblock The aflow standard for high-throughput materials science
  calculations.
\newblock {\em Computational Materials Science}, 108:233--238, 2015.

\bibitem{hachmann2011harvard}
Johannes Hachmann, Roberto Olivares-Amaya, Sule Atahan-Evrenk, Carlos
  Amador-Bedolla, Roel~S S{\'a}nchez-Carrera, Aryeh Gold-Parker, Leslie Vogt,
  Anna~M Brockway, and Al{\'a}n Aspuru-Guzik.
\newblock The harvard clean energy project: large-scale computational screening
  and design of organic photovoltaics on the world community grid.
\newblock {\em The Journal of Physical Chemistry Letters}, 2(17):2241--2251,
  2011.

\bibitem{jain2013commentary}
Anubhav Jain, Shyue~Ping Ong, Geoffroy Hautier, Wei Chen, William~Davidson
  Richards, Stephen Dacek, Shreyas Cholia, Dan Gunter, David Skinner, Gerbrand
  Ceder, et~al.
\newblock Commentary: The materials project: A materials genome approach to
  accelerating materials innovation.
\newblock {\em APL materials}, 1(1):011002, 2013.

\bibitem{dan2020computational}
Yabo Dan, Rongzhi Dong, Zhuo Cao, Xiang Li, Chengcheng Niu, Shaobo Li, and
  Jianjun Hu.
\newblock Computational prediction of critical temperatures of superconductors
  based on convolutional gradient boosting decision trees.
\newblock {\em IEEE Access}, 8:57868--57878, 2020.

\bibitem{jha2021enabling}
Dipendra Jha, Vishu Gupta, Logan Ward, Zijiang Yang, Christopher Wolverton, Ian
  Foster, Wei-keng Liao, Alok Choudhary, and Ankit Agrawal.
\newblock Enabling deeper learning on big data for materials informatics
  applications.
\newblock {\em Scientific reports}, 11(1):1--12, 2021.

\bibitem{xie2018crystal}
Tian Xie and Jeffrey~C Grossman.
\newblock Crystal graph convolutional neural networks for an accurate and
  interpretable prediction of material properties.
\newblock {\em Physical review letters}, 120(14):145301, 2018.

\bibitem{omee2022scalable}
Sadman~Sadeed Omee, Steph-Yves Louis, Nihang Fu, Lai Wei, Sourin Dey, Rongzhi
  Dong, Qinyang Li, and Jianjun Hu.
\newblock Scalable deeper graph neural networks for high-performance materials
  property prediction.
\newblock {\em Patterns}, 3(5):100491, 2022.

\bibitem{wang2019machine}
Ruirui Wang, Shuming Zeng, Xinming Wang, and Jun Ni.
\newblock Machine learning for hierarchical prediction of elastic properties in
  fe-cr-al system.
\newblock {\em Computational Materials Science}, 166:119--123, 2019.

\bibitem{zhao2020predicting}
Yong Zhao, Kunpeng Yuan, Yinqiao Liu, Steph-Yves Louis, Ming Hu, and Jianjun
  Hu.
\newblock Predicting elastic properties of materials from electronic charge
  density using 3d deep convolutional neural networks.
\newblock {\em The Journal of Physical Chemistry C}, 124(31):17262--17273,
  2020.

\bibitem{mazhnik2020application}
Efim Mazhnik and Artem~R Oganov.
\newblock Application of machine learning methods for predicting new superhard
  materials.
\newblock {\em Journal of Applied Physics}, 128(7):075102, 2020.

\bibitem{al2021high}
Mohammed Al-Fahdi, Tao Ouyang, and Ming Hu.
\newblock High-throughput computation of novel ternary b--c--n structures and
  carbon allotropes with electronic-level insights into superhard materials
  from machine learning.
\newblock {\em Journal of Materials Chemistry A}, 9(48):27596--27614, 2021.

\bibitem{hu2022piezoelectric}
Jeffrey Hu and Yuqi Song.
\newblock Piezoelectric modulus prediction using machine learning and graph
  neural networks.
\newblock {\em Chemical Physics Letters}, 791:139359, 2022.

\bibitem{han2021machine}
Guangshuai Han, Yixuan Sun, Yining Feng, Guang Lin, and Na~Lu.
\newblock Machine learning regression guided thermoelectric materials
  discovery--a review.
\newblock {\em ES Materials \& Manufacturing}, 14:20--35, 2021.

\bibitem{loftis2020lattice}
Christian Loftis, Kunpeng Yuan, Yong Zhao, Ming Hu, and Jianjun Hu.
\newblock Lattice thermal conductivity prediction using symbolic regression and
  machine learning.
\newblock {\em The Journal of Physical Chemistry A}, 125(1):435--450, 2020.

\bibitem{ouyang2021machine}
Yulou Ouyang, Cuiqian Yu, Gang Yan, and Jie Chen.
\newblock Machine learning approach for the prediction and optimization of
  thermal transport properties.
\newblock {\em Frontiers of Physics}, 16(4):1--16, 2021.

\bibitem{louis2022accurate}
Steph-Yves Louis, Edirisuriya M~Dilanga Siriwardane, Rajendra~P Joshi,
  Sadman~Sadeed Omee, Neeraj Kumar, and Jianjun Hu.
\newblock Accurate prediction of voltage of battery electrode materials using
  attention-based graph neural networks.
\newblock {\em ACS Applied Materials \& Interfaces}, 2022.

\bibitem{belli2021strong}
Francesco Belli, Trinidad Novoa, Julia Contreras-Garc{\'\i}a, and Ion Errea.
\newblock Strong correlation between electronic bonding network and critical
  temperature in hydrogen-based superconductors.
\newblock {\em Nature communications}, 12(1):1--11, 2021.

\bibitem{xie2022machine}
SR~Xie, Y~Quan, AC~Hire, B~Deng, JM~DeStefano, I~Salinas, US~Shah,
  L~Fanfarillo, J~Lim, J~Kim, et~al.
\newblock Machine learning of superconducting critical temperature from
  eliashberg theory.
\newblock {\em npj Computational Materials}, 8(1):1--8, 2022.

\bibitem{dong2022deepxrd}
Rongzhi Dong, Yong Zhao, Yuqi Song, Nihang Fu, Sadman~Sadeed Omee, Sourin Dey,
  Qinyang Li, Lai Wei, and Jianjun Hu.
\newblock Deepxrd, a deep learning model for predicting of xrd spectrum from
  materials composition.
\newblock {\em arXiv preprint arXiv:2203.14326}, 2022.

\bibitem{gupta2021cross}
Vishu Gupta, Kamal Choudhary, Francesca Tavazza, Carelyn Campbell, Wei-keng
  Liao, Alok Choudhary, and Ankit Agrawal.
\newblock Cross-property deep transfer learning framework for enhanced
  predictive analytics on small materials data.
\newblock {\em Nature communications}, 12(1):1--10, 2021.

\bibitem{abdar2021review}
Moloud Abdar, Farhad Pourpanah, Sadiq Hussain, Dana Rezazadegan, Li~Liu,
  Mohammad Ghavamzadeh, Paul Fieguth, Xiaochun Cao, Abbas Khosravi, U~Rajendra
  Acharya, et~al.
\newblock A review of uncertainty quantification in deep learning: Techniques,
  applications and challenges.
\newblock {\em Information Fusion}, 76:243--297, 2021.

\bibitem{nigam2021assigning}
AkshatKumar Nigam, Robert Pollice, Matthew~FD Hurley, Riley~J Hickman, Matteo
  Aldeghi, Naruki Yoshikawa, Seyone Chithrananda, Vincent~A Voelz, and Al{\'a}n
  Aspuru-Guzik.
\newblock Assigning confidence to molecular property prediction.
\newblock {\em Expert opinion on drug discovery}, 16(9):1009--1023, 2021.

\bibitem{hirschfeld2020uncertainty}
Lior Hirschfeld, Kyle Swanson, Kevin Yang, Regina Barzilay, and Connor~W Coley.
\newblock Uncertainty quantification using neural networks for molecular
  property prediction.
\newblock {\em Journal of Chemical Information and Modeling}, 60(8):3770--3780,
  2020.

\bibitem{xin2021active}
Rui Xin, Edirisuriya~MD Siriwardane, Yuqi Song, Yong Zhao, Steph-Yves Louis,
  Alireza Nasiri, and Jianjun Hu.
\newblock Active-learning-based generative design for the discovery of
  wide-band-gap materials.
\newblock {\em The Journal of Physical Chemistry C}, 125(29):16118--16128,
  2021.

\bibitem{tran2020methods}
Kevin Tran, Willie Neiswanger, Junwoong Yoon, Qingyang Zhang, Eric Xing, and
  Zachary~W Ulissi.
\newblock Methods for comparing uncertainty quantifications for material
  property predictions.
\newblock {\em Machine Learning: Science and Technology}, 1(2):025006, 2020.

\bibitem{pmlr-v37-blundell15}
Charles Blundell, Julien Cornebise, Koray Kavukcuoglu, and Daan Wierstra.
\newblock Weight uncertainty in neural network.
\newblock In Francis Bach and David Blei, editors, {\em Proceedings of the 32nd
  International Conference on Machine Learning}, volume~37 of {\em Proceedings
  of Machine Learning Research}, pages 1613--1622, Lille, France, 07--09 Jul
  2015. PMLR.

\bibitem{NEURIPS2018_a981f2b7}
Murat Sensoy, Lance Kaplan, and Melih Kandemir.
\newblock Evidential deep learning to quantify classification uncertainty.
\newblock In S.~Bengio, H.~Wallach, H.~Larochelle, K.~Grauman, N.~Cesa-Bianchi,
  and R.~Garnett, editors, {\em Advances in Neural Information Processing
  Systems}, volume~31. Curran Associates, Inc., 2018.

\bibitem{amini2020deep}
Alexander Amini, Wilko Schwarting, Ava Soleimany, and Daniela Rus.
\newblock Deep evidential regression.
\newblock {\em Advances in Neural Information Processing Systems},
  33:14927--14937, 2020.

\bibitem{zhang_conf_2021}
Jin Zhang, Ulf Norinder, and Fredrik Svensson.
\newblock Deep learning-based conformal prediction of toxicity.
\newblock {\em Journal of Chemical Information and Modeling}, 61(6):2648--2657,
  2021.

\bibitem{gentle_conf}
Anastasios~N Angelopoulos and Stephen Bates.
\newblock A gentle introduction to conformal prediction and distribution-free
  uncertainty quantification.
\newblock {\em arXiv preprint arXiv:2107.07511}, 2021.

\bibitem{svensson_conf_2018}
Fredrik Svensson, Avid~M. Afzal, Ulf Norinder, and Andreas Bender.
\newblock Maximizing gain in high-throughput screening using conformal
  prediction.
\newblock {\em Journal of Cheminformatics}, 10(1):7, 2018.

\bibitem{diet_ens}
Thomas~G. Dietterich.
\newblock Ensemble methods in machine learning.
\newblock In {\em Multiple Classifier Systems}, International Workshop on
  Multiple Classifier Systems, pages 1--15. Springer Berlin Heidelberg, 2000.

\bibitem{lakshminarayanan2017simple}
Balaji Lakshminarayanan, Alexander Pritzel, and Charles Blundell.
\newblock Simple and scalable predictive uncertainty estimation using deep
  ensembles.
\newblock {\em Advances in neural information processing systems}, 30, 2017.

\bibitem{leibig2017leveraging}
Christian Leibig, Vaneeda Allken, Murat~Se{\c{c}}kin Ayhan, Philipp Berens, and
  Siegfried Wahl.
\newblock Leveraging uncertainty information from deep neural networks for
  disease detection.
\newblock {\em Scientific reports}, 7(1):1--14, 2017.

\bibitem{cortes2018deep}
Isidro Cort{\'e}s-Ciriano and Andreas Bender.
\newblock Deep confidence: a computationally efficient framework for
  calculating reliable prediction errors for deep neural networks.
\newblock {\em Journal of chemical information and modeling}, 59(3):1269--1281,
  2018.

\bibitem{wen2020uncertainty}
Mingjian Wen and Ellad~B Tadmor.
\newblock Uncertainty quantification in molecular simulations with dropout
  neural network potentials.
\newblock {\em npj Computational Materials}, 6(1):1--10, 2020.

\bibitem{korolev2022universal}
Vadim Korolev, Iurii Nevolin, and Pavel Protsenko.
\newblock A universal similarity based approach for predictive uncertainty
  quantification in materials science.
\newblock {\em Scientific Reports}, 12(1):1--10, 2022.

\bibitem{scarselli2008graph}
Franco Scarselli, Marco Gori, Ah~Chung Tsoi, Markus Hagenbuchner, and Gabriele
  Monfardini.
\newblock The graph neural network model.
\newblock {\em IEEE transactions on neural networks}, 20(1):61--80, 2008.

\bibitem{gnn_base}
Franco Scarselli, Marco Gori, Ah~Chung Tsoi, Markus Hagenbuchner, and Gabriele
  Monfardini.
\newblock The graph neural network model.
\newblock {\em IEEE Transactions on Neural Networks}, 20(1):61--80, 2009.

\bibitem{matDL_pap}
Victor Fung, Jiaxin Zhang, Eric Juarez, and Bobby~G Sumpter.
\newblock Benchmarking graph neural networks for materials chemistry.
\newblock {\em npj Computational Materials}, 7(1):1--8, 2021.

\bibitem{schutt2017schnet}
Kristof Sch{\"u}tt, Pieter-Jan Kindermans, Huziel~Enoc Sauceda~Felix, Stefan
  Chmiela, Alexandre Tkatchenko, and Klaus-Robert M{\"u}ller.
\newblock Schnet: A continuous-filter convolutional neural network for modeling
  quantum interactions.
\newblock {\em Advances in neural information processing systems}, 30, 2017.

\bibitem{schwarzer2019learning}
Max Schwarzer, Bryce Rogan, Yadong Ruan, Zhengming Song, Diana~Y Lee, Allon~G
  Percus, Viet~T Chau, Bryan~A Moore, Esteban Rougier, Hari~S Viswanathan,
  et~al.
\newblock Learning to fail: Predicting fracture evolution in brittle material
  models using recurrent graph convolutional neural networks.
\newblock {\em Computational Materials Science}, 162:322--332, 2019.

\bibitem{chen2019graph}
Chi Chen, Weike Ye, Yunxing Zuo, Chen Zheng, and Shyue~Ping Ong.
\newblock Graph networks as a universal machine learning framework for
  molecules and crystals.
\newblock {\em Chemistry of Materials}, 31(9):3564--3572, 2019.

\bibitem{wieder2020compact}
Oliver Wieder, Stefan Kohlbacher, M{\'e}laine Kuenemann, Arthur Garon, Pierre
  Ducrot, Thomas Seidel, and Thierry Langer.
\newblock A compact review of molecular property prediction with graph neural
  networks.
\newblock {\em Drug Discovery Today: Technologies}, 37:1--12, 2020.

\bibitem{wugnnsurvey}
Zonghan Wu, Shirui Pan, Fengwen Chen, Guodong Long, Chengqi Zhang, and
  Philip~S. Yu.
\newblock A comprehensive survey on graph neural networks.
\newblock {\em IEEE Transactions on Neural Networks and Learning Systems},
  32(1):4--24, 2021.

\bibitem{eklund_conf_2015}
Martin Eklund, Ulf Norinder, Scott Boyer, and Lars Carlsson.
\newblock The application of conformal prediction to the drug discovery
  process.
\newblock {\em Annals of Mathematics and Artificial Intelligence},
  74(1):117--132, 2015.

\bibitem{amini_2021}
Ava~P. Soleimany, Alexander Amini, Samuel Goldman, Daniela Rus, Sangeeta~N.
  Bhatia, and Connor~W. Coley.
\newblock Evidential deep learning for guided molecular property prediction and
  discovery.
\newblock {\em ACS Central Science}, 7(8):1356--1367, 08 2021.

\bibitem{delta_metric}
Vadim Korolev, Iurii Nevolin, and Pavel Protsenko.
\newblock A universal similarity based approach for predictive uncertainty
  quantification in materials science.
\newblock {\em Scientific Reports}, 12(1):14931, 2022.

\bibitem{bartok2013representing}
Albert~P Bart{\'o}k, Risi Kondor, and G{\'a}bor Cs{\'a}nyi.
\newblock On representing chemical environments.
\newblock {\em Physical Review B}, 87(18):184115, 2013.

\bibitem{janet2019quantitative}
Jon~Paul Janet, Chenru Duan, Tzuhsiung Yang, Aditya Nandy, and Heather~J Kulik.
\newblock A quantitative uncertainty metric controls error in neural
  network-driven chemical discovery.
\newblock {\em Chemical science}, 10(34):7913--7922, 2019.

\bibitem{mamun2019high}
Osman Mamun, Kirsten~T Winther, Jacob~R Boes, and Thomas Bligaard.
\newblock High-throughput calculations of catalytic properties of bimetallic
  alloy surfaces.
\newblock {\em Scientific data}, 6(1):1--9, 2019.

\bibitem{fung2017exploring}
Victor Fung and De-en Jiang.
\newblock Exploring structural diversity and fluxionality of pt n (n= 10--13)
  clusters from first-principles.
\newblock {\em The Journal of Physical Chemistry C}, 121(20):10796--10802,
  2017.

\bibitem{rosen2021machine}
Andrew~S Rosen, Shaelyn~M Iyer, Debmalya Ray, Zhenpeng Yao, Alan Aspuru-Guzik,
  Laura Gagliardi, Justin~M Notestein, and Randall~Q Snurr.
\newblock Machine learning the quantum-chemical properties of metal--organic
  frameworks for accelerated materials discovery.
\newblock {\em Matter}, 4(5):1578--1597, 2021.

\end{thebibliography}

\end{document}